# Statistical distances between countries and cluster structures in EU area according to macroeconomic indices fluctuations

Mircea Gligor[*] [†] Marcel Ausloos[†]

[*] National College "Roman Voda", Romania, mgligor@ulg.ac.be
[†] University of Liège, Euroland, Marcel.Ausloos@ulg.ac.be

# Statistical distances between countries and cluster structures in EU area according to macroeconomic indices fluctuations*

Mircea Gligor and Marcel Ausloos

**Abstract**

The paper applies some recent developments of network analysis in order to perform a comparative study of EU countries in relation with the fluctuations of some macroeconomic indicators. The statistical distances between countries, calculated for various moving average time windows, are mapped into the ultrametric subdominant space as in classical Minimal Spanning Tree methods. The novelty consists in applying the so-called Moving Average Minimal Length Path (MAMLP) algorithm, which allows a decoupling of fluctuations with respect to the mass center of the system from the movement of the mass center itself. The present analysis pertains to the Gross Domestic Product and some of its sources, namely the Final Consumption Expenditure, Gross Capital Formation and Net Exports growth rates. The target group of countries is composed of 15 EU countries, data taken between 1972 and 2004, i.e., before the last wave of integration. The method allows us to search for a cluster-like structure when derived both from the hierarchical organization of countries and from their relative movement inside the hierarchy. It is found that the strongly correlated countries with respect to GDP fluctuations can be partitioned into stable clusters. Some of the highly correlated countries, with respect to GDP fluctuations, display strong correlations also in the Final Consumption Expenditure, while others are strongly correlated in Gross Capital Formation. On the other hand, one notices the similitude of the classifications regarding GDP and Net Exports fluctuations as concerns the squared sum of the correlation coefficients (so called country "sensitivity"). We find that, as for social networks, there is a positive correlation between the node sensitivities i.e. the high connected countries commonly tend to connect with other well connected ones. The final structure proves to be stable against the fluctuations induced by the moving time window over the scanned time interval.

**KEYWORDS**: fluctuations, correlations, moving average, minimal length path, clustering

* Gligor: mgligor@ulg.ac.be     Ausloos: Marcel.Ausloos@ulg.ac.be

# 1 Introduction

The political and social evolution of EU countries during the last decade points out to both cohesion and centrifugal tendencies that are expected to exert some influence on their macroeconomic evolution. Various and quite different positions have been adopted in relation with political issues (the war of Iraq, the EU extension, the EU Constitution), but also in relation to particular economic issues, *e.g.* the common currency, the interest rates, environmental problems. To analyse the impact of each one of these factors upon the whole EU system as well as the role that each country plays are rather difficult tasks, taking into account that two or more countries may adopt the same position with respect to a problem, and opposite positions with respect to another one.

On the other hand, in the course of nature, the European countries can be considered as interacting agents of a complex system; it is expected that the economic fluctuations which occur in one country influence in turns the fluctuations in the other ones, especially those having the strongest economic connections with the former. This aspect leads to imagine a hierarchical organisation and a cluster-like structure of EU countries, as found in many natural and social complex systems investigated so far. The goal of the present paper is not to analyse the impact of major events or political decisions on this system, but to investigate the strongest correlations and anti-correlations in the dynamical evolution of some macro-economic indicators in an empirical way, focusing on those which are the most closely related to development *i.e.* the Gross Domestic Product (GDP) and GDP/capita, as well as the ones which are usually considered for GDP estimation: the Final Consumption Expenditure (FCE), the Gross Capital Formation (GCF) and the Net Exports (NEX = exports – imports of goods and services).

There is a large literature devoted to the distribution of shocks to GDP growth rates and their standard deviation scaling properties (*e.g.* Canning et al (1998), and references therein), and also numerous papers analyzing the cross-correlations between OECD countries from various points of view. One could mention here Cooper and Haltiwanger, (1996) where emphasis is placed on calibrating models to explain the variances and covariances of macroeconomic variables, Kraay and Ventura (2002), which sets off the remarkable degree of synchronization of the business cycles in OECD countries and Ludwig and Sløk (2004), which analyze the relationship between stock prices, house prices and consumption using data for 16 OECD countries.

The present paper tries to set the problem of cross-correlations into the general frameworks of classification trees and complex systems analysis. It is well known that a general question facing researchers in many areas of inquiry is how to organize observed data into meaningful structures, that is, to develop taxonomies. In this sense, cluster analysis is an exploratory data analysis tool which aims at sorting different objects into groups in a way that the degree of association between two objects is maximal if they belong to the same group and minimal otherwise. The term “cluster analysis” (first used by Tryon, 1939) refers to a number of different algorithms and methods for grouping objects of similar kinds into respective categories. Unlike many other statistical procedures, cluster analysis methods are mostly used when we do not have any a priori hypothesis, but are still in a exploratory research phase. In general, whenever one needs to classify a large amount of information into manageable meaningful sequences, cluster analysis is of great utility.

As we intend to build a cluster-like structure based on the strongest correlations and anti-correlations between time series, our approach has some common points with other classification tree methods of interest in statistics. Long ago, methods as CHAID (Chi-squared Automatic Interaction Detector) proposed by Kass (1980), the classical C&R Trees (Classification and Regression Trees) Algorithm (Breiman et al., 1984) and other tree classification techniques have been discussed. They are known to have a number of advantages over many other techniques. In most cases, the interpretation of results summarized as on a tree is very simple. This simplicity is useful not only for purposes of rapid classification of new observations, but can also often yield a simple “model” for explaining why observations are ordered or predicted in a particular manner. On the other hand, the final results of using tree methods for classification or regression can be summarized in a series of (usually few) logical if-then conditions (tree nodes). Therefore, there is no implicit assumption on the underlying relationships between the predictor variables. Thus, tree methods are particularly well suited for data mining tasks, when there is often little a priori knowledge nor any coherent set of theories or predictions regarding which variables are related and how.

As a matter of fact, the present paper deals with a classification-type problem that is to predict values of a categorical dependent variable (class, group membership, etc.) from a predictor variable which is - in our approach - the correlation coefficient.

A tree clustering method uses the dissimilarities (similarities) measured as distances between objects when forming the clusters. Therefore, in tree-like classifications, the first problem is to choose an adequate distance measure in order to place progressively greater weight on objects (say series $\{x_i\}$ and $\{y_i\}$) that are further apart.

Various definitions are proposed in the statistics literature so far. We recall here only those of common use, as the Euclidean distance:

$$d(x, y) = \left( \sum_i (x_i - y_i)^2 \right)^{1/2} \qquad (1.1)$$

and the City-block (Manhattan) distance:

$$d(x,y) = \sum_i | x_i - y_i | \quad (1.2)$$

The first method has a few advantages (e.g., the distance between any two objects is not affected by the addition of new objects in the analysis, which may be outliers). In (1.2) the effect of outliers is dampened since they are not squared. The distance (1.1) can be generalized as a power distance:

$$d(x,y) = \left( \sum_i (x_i - y_i)^p \right)^{1/r} \quad (1.3)$$

where $p$ and $r$ are user-defined parameters, or as a correlation (statistical) distance:

$$d(x,y) = \left[ 2(1 - C(x,y)) \right]^{1/2} \quad (1.4)$$

where the $C$ is the correlation coefficient:

$$C(x,y) = \frac{< x_i y_i > - < x_i >< y_i >}{\sqrt{< x_i^2 - < x_i >^2 >< y_i^2 - < y_i >^2 >}} \quad (1.5)$$

As we aim to build a hierarchical structure starting from the correlations between time series, in the present work we opted for (1.4), though we recognize that other choices could be possible.

The method used here below, namely the moving-average-minimal-length-path (MAMLP) is described in Section 2**,** with other several related techniques. In essence, MAMLP was derived by applying the minimal-length-path-to-average classification to various moving time windows. In other words, as a first step, for each time window a hierarchy of countries was found taking their minimal path distance on average; thereafter, in a second step the strongest correlations and anti-correlations between the movements of countries inside the hierarchy were investigated.

In Section 2.4 the macroeconomic indicators whose fluctuations* have been considered are briefly presented, and also, the data sources which are used. The target group of countries is composed of 15 EU countries; the data refers to years between 1972 and 2004 (for the 10 years size time window analysis) and between 1994 and 2004 (for the 5 years size time window analysis case), that is before the last wave of EU extension.

The results are presented in Section 3. This section has a tri-partite structure, grouping the results in relation with the multiple aims of our investigation: first, the relevant role of the time window size is pointed out by studying GDP/capita in two moving time windows of 10 and 5 years sizes respectively; secondly, GDP is investigated in a moving time window

* The term “fluctuations” refers to the annual rates of growth of the considered indicators.

of 5 years, and the MAMLP method is applied to find the strongest correlations and anti-correlations between countries, which result in a cluster-like structure; thirdly, the same method is applied to the other three indicators (FCE, GCF and NEX), which are usually considered as basic ingredients in the GDP estimations.

Conclusions are found in Section 4. A statistical test of robustness, – the shuffled data analysis – is done in Appendix.

# 2 Theoretical and methodological framework

## 2.1 The Minimal Spanning Tree (MST)

First used by Mantegna (1999) in the econophysics field, the MST can be seen as a modern extension of the Horizontal- (or Vertical-) Hierarchical-Tree-Plot – an older clustering method well known for its large applicability in medicine, psychiatry and archeology (Hartigan, 1975). The essential additional ingredients of MST consist in the use of the ultrametric subdominant space and of the ultrametric distance between objects.

In the above cited paper Mantegna analyzed the correlations between the 30 stocks used to calculate the Dow Jones Industrial Average, on one hand, and correlations between the companies used to calculate the S&P 500 index, on the other hand, for the July 1989 to October 1995 time span in both cases. Only the companies which were present in the S&P 500 index during the whole time interval were considered *i.e.,* 443 stocks. The correlation coefficients $C(x, y)$ of the returns for all possible pairs of stocks were computed. A metric distance between two stocks was defined by:

$$d(x, y) = 1 - [C(x, y)]^2$$

These distances become the elements of the distance matrix $D$, from which he determined a topological arrangement between the different stocks. His study could also give some empirical evidence about the existence and nature of common economic factors which drive the time evolution of stock prices. He determined the minimum spanning tree (MST) connecting the different indices, and thereafter, the subdominant ultrametric structure and hierarchical organization of the indices. In fact, the elements $\hat{d}_{ij}$ of the ultrametric distance matrix $\hat{D}$ are determined from the MST, where $\hat{d}_{ij}$ is the maximum Euclidian distance detected by moving from $i$ to $j$ by single steps through the path connecting $i$ to $j$ in the MST. Studying the MST and the hierarchical organization of the stocks defining the Dow Jones industrial average, Mantegna showed that the stocks can be divided into three groups. Carrying the same analysis for the stocks belonging to the S&P500, Mantegna (1999) obtained clusters of the stocks according to the industry they belong to.

## 2.2 Local Minimal Spanning Tree and the Minimal Length Path algorithms

Unlike the high frequency financial data series, the macroeconomic time series are too short and noisy. Most macroeconomic data have a yearly or at most quarterly frequency. When such time series have been produced for a very long period, there is usually strong evidence against stationarity. A way for investigating such time series can only be by moving a constant size time window with a constant step so that the whole time interval is scanned.

Moreover, when too small differences between the linkage distances arise, MST seems to be not unique. This aspect can lead either to a maximal dispersed structure (each object is in a class by itself) or, contrarily, to a high clustered structure in which all objects are joined together. This aspect was signaled by many authors (*e.g.* Bouchaud and Potters, 2000).

Some alternative ways for constructing the hierarchy, better adapted to the low frequency time series have been recently proposed. The Local Minimum Spanning Tree (LMST) is a modification of the MST algorithm under the constraint that the initial pair of nodes (the root) of the tree is the pair with the strongest correlation. Correlation chains have been investigated in the context of the most developed countries clustering in two forms: unidirectional and bidirectional minimum length chains (UMLP and BMLP respectively) (Miskiewicz and Ausloos, 2005). UMLP and BMLP algorithms are simplifications for LMST, where the closest neighbouring countries are attached at the end of a chain. In the case of the unidirectional chain the initial node is an arbitrary chosen country. Therefore in the case of UMLP the chain is expanded in one direction only, whereas in the bidirectional case countries might be attached at one of both ends depending on the distance value. These authors also underlined some arbitrariness in the root of the tree for comparing results, and considered that an a priori more common root, like the sum of the data, called the “All” country, from which to let the tree grow was permitting a better comparison.

## 2.3 The Moving-Average Minimal Length Path (MAMLP) Method

The problem that MST cannot be built in a unique way becomes even more important when we try to construct a cluster hierarchy for each position of a moving time window. The hierarchical structure proved to be not robust against fluctuations induced by a moving time window. Simply, if the statistical distances between pairs A-B and C-D belonging to different clusters are small, it is quite likely to find at the next step A-C and B-D as pairs in other different clusters.

In the MAMLP method described here below we propose to construct the hierarchy also starting from a virtual 'average' agent. In fact, the method of decoupling the movement of the mass center and the movement of independent parts is quite of common use in physics and statistics.

The method is developed in the following steps:

i. An 'AVERAGE' agent (AV) is virtually included into the system;
ii. The statistical distance matrix is constructed, and thereafter, the elements are set into increasing order (i.e. the decreasing order of correlations);
iii. The hierarchy is constructed, connecting each agent by its minimal length path to AV. Its minimal distance to AV is associated to each agent.
iv. The procedure is repeated by moving a given and constant time window over the investigated time span. The statistical properties of the datasets are investigated.
v. The agents are sorted either through their mean position in the hierarchy during the period investigated, or through their movement inside the hierarchy. In the former case, the distances to the AV are normalized to the spanning range of the hierarchy. In the latter version, a new correlation matrix between country distances to their own mean is constructed.

## 2.4 Data sources

Macroeconomic indicators freely taken from the web form the data base used for the present investigations. We consider: (a) Annual percentage growth rate of GDP and GDP/capita at market prices based on constant local currency. Aggregates are based on constant 2000 U.S. dollars. GDP is the sum of gross value added by all resident producers in the economy plus any product taxes and minus any subsidies not included in the value of the products. It is calculated without making deductions for depreciation of fabricated assets or for depletion and degradation of natural resources. (b) Average annual growth of Final Consumption Expenditure (FCE) based on constant local currency. Aggregates are based on constant 2000 U.S. dollars. Final Consumption Expenditure (formerly total consumption) is the sum of household final consumption expenditure (formerly private consumption) and general government final consumption expenditure (formerly general government consumption); (c) Annual growth rate of Gross Capital Formation (GCF) based on constant local currency. Aggregates are based on constant 2000 U.S. dollars. Gross Capital Formation (formerly gross domestic investment) consists of outlays on additions to the fixed assets of the economy plus net changes in the level of inventories. Fixed assets include land improvements (fences, ditches, drains, and so on); plant, machinery, and equipment purchases; and the construction of roads, railways, and the like, including schools, offices, hospitals, private residential dwellings, and commercial and industrial buildings. Inventories are stocks of goods held by firms to meet temporary or unexpected fluctuations in production or sales, and "work in progress." According to the 1993 SNA, net acquisitions of valuables are also considered; (d) Annual growth rate of net exports (NEX)**,** calculated as the difference between the annual growths rates of exports and imports of goods and services based on constant local currency. Aggregates are based on constant 2000 U.S. dollars. Exports of goods and services represent the value of all goods and other market services provided to the rest of the world. Imports of goods and services represent the value of all goods and other market services received from the rest of the world. They include the value of merchandise, freight, insurance, transport, travel, royalties, license fees, and other services, such as communication, construction, financial, information, business, personal, and government services. They exclude labor and property income (formerly called factor services) as well as transfer payments. including here the value of merchandise, freight, insurance, transport, travel, royalties, license fees, and other services, such as communication, construction, financial, information, business, personal, and government services.

The main source used for all the above indicators annual rates of growth taken between 1972 and 2004 is here below the World Bank database:
http://devdata.worldbank.org/query/default.htm.

In addition to the above mentioned data bank, for comparison aims, we also used the data supplied by:
http://www.economicswebinstitute.org/concepts.htm (1986-2000);
http://www.oecd.org/about/0,2337,en_2649_201185_1_1_1_1_1,00.html (2003-2004).

We abbreviate the countries according to The Roots Web Surname List (RSL) which uses 3 letters standardized abbreviations to designate countries and other regional locations (http://helpdesk.rootsweb.com/codes/). Inside the tables, for spacing reasons we use the countries two letters abbreviation (http://www.iso.org/iso/en/prods-services/iso3166ma/02iso-3166-code-lists/list-en1.html).

Finally, we assume that the database for the 15 EU countries satisfies the requirements of international comparison, *i.e.* characteristicity, base-country invariance, transitivity, additivity and commensurability. The role of the concrete aggregation method – by averaging of bilateral indices or by use of average international prices – is discounted in some extent by using the relative variations of the investigated quantities.

# 3 Results

## 3.1 The mean statistical distance between EU countries in various time window sizes

GDP/capita data is first investigated with a fixed $T$ = 10 years moving time window size, and the statistical distance matrix $D$ thereby constructed, taking into account $N$ = 15 countries, namely AUT, BEL, DEU, DNK, ESP, FIN, FRA, GBR, GRC, IRL, ITA, LUX, NLD, PRT and SWE. The mean distance between the countries $<d>$ is calculated by averaging all the statistical distances from $D$, for each time interval:

$$< d >_{(t,t+T)} = \frac{1}{2} \sum_{\substack{i,j=1 \\ i \neq j}}^{N} d_{ij} \tag{3.1}$$

In order to identify the trend of $<d>$, we use the reduced mean statistical distance, defined as:

$$< \tilde{d} >_{(t,t+T)} = \frac{1}{\sigma} < d >_{(t,t+T)} \tag{3.2}$$

where:

$$\sigma \equiv \sigma(t,T) = \left[ \frac{1}{N} \sum_{\substack{i,j=1 \\ i > j}}^{N} [d_{ij} - < d >_{(t,t+T)}]^2 \right] \tag{3.3}$$

is the standard deviation of the dataset.

In Figure 1 the reduced mean statistical distance is plotted taking into account all 15 EU-countries, between 1972 and 2004, by moving the 10 years time window by a year time step. For simplicity, the interval notation is abbreviated at the last two digits of the first and last year of the window, and each data point is arbitrarily centered in the middle of the interval.

The time evolution of $< \tilde{d} >$ sets off a succession of abrupt increases ("shocks") followed by decreases ("relaxations"). Such phenomenon, occurred in the time interval 1986-2004, is separately plotted in Figure 2. The variable $x$ of the fit function (in the inset) represents the order number of the point. The time variation of $< \tilde{d} >$ displays an unexpected abrupt jump when going from 1991-2000 to 1992-2001, followed by a decay well fitted by an exponential (see inset). If the exponential decay is written as: $< \tilde{d} >= (const)\exp(-x/\tau)$, then $\tau$ is the so called "relaxation time" of the process. Here it is about 12.5 years. The abrupt jump of $< \tilde{d} >$ is seen together with some similar anomaly in other statistical properties of the $\{d_{ij}\}$ datasets, as the variance, kurtosis and skewness (see Figure 3). Suspecting an effect due to Germany reunification, the data has been reanalyzed and is also shown on the same figure, but for only 14 countries (removing DEU – Figure 2), - but the anomalies remain.

In the next step of investigation, the second branch, *i.e.* the time interval 1994-2004, is scanned with a shorter 5 years moving time window. A monotonic decreasing trend is again easily noticeable in Figure 4, corresponding to a relaxation time of the same order of magnitude, i.e., $\tau \sim$ 8-10 years.

In view of this time window effect**,** it seems reasonable to study the statistical distances between countries using GDP, FCE and GCF annual growth rates for the same (short) 5 years moving time window, for the data taken from 1994 to 2004[*]. It is seen that the reduced mean distance among the EU-15 countries, as plotted in Figure 5, follows the same decreasing trend as in Figure 4 for the GDP/capita, indicating a remarkable degree of similarity between the after-shock responses of the system with respect to GDP and GCF fluctuations (the same relaxation time $\tau \sim$ 8-10 years is found as in the case of GDP/capita). The relaxation time is $\tau$ > 10 years for FCE fluctuations. We recall here that the term "fluctuations" refers, as above, to the annual rates of growth of the considered indicators (see data in insets).

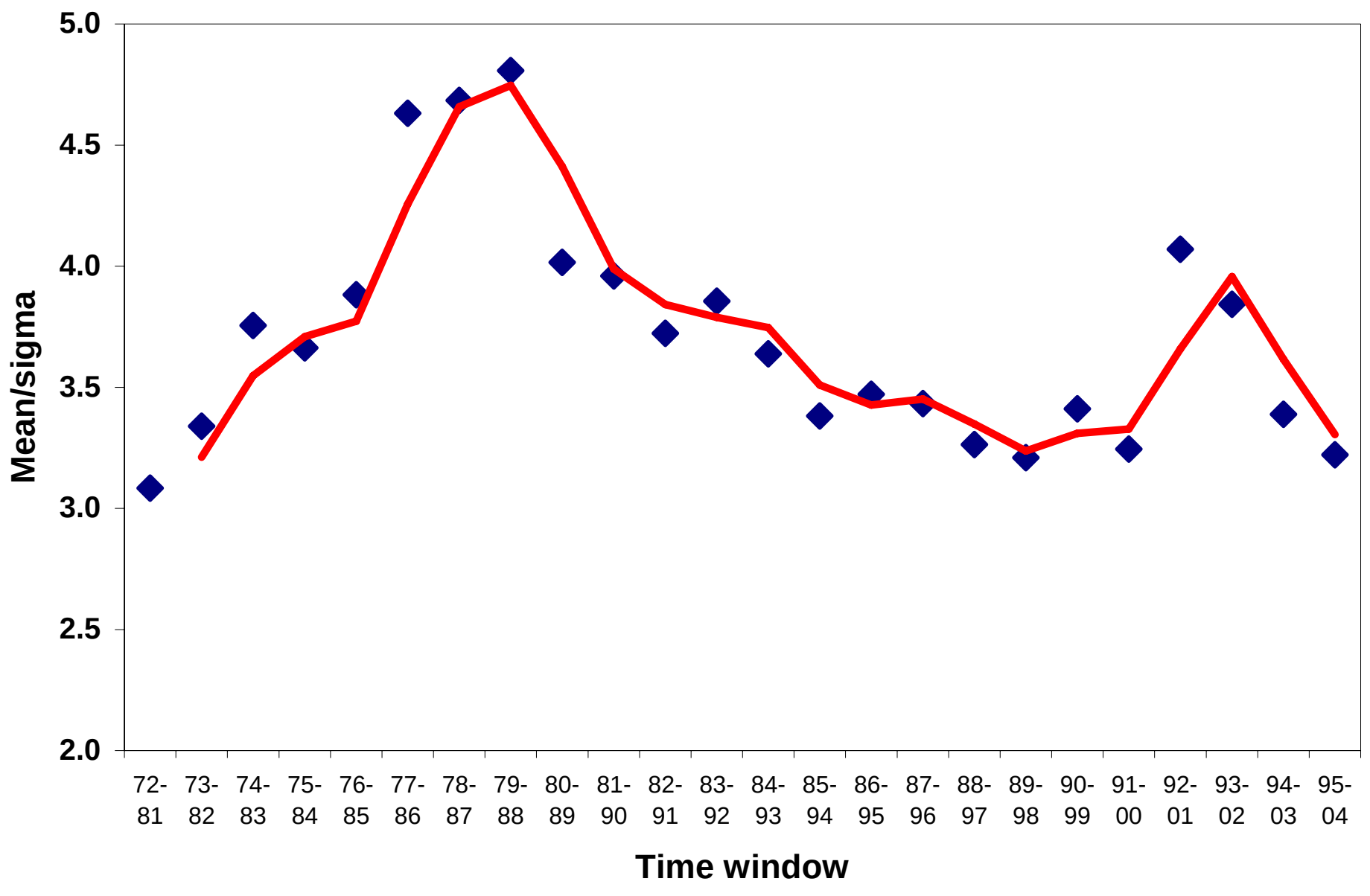

Figure 1: The GDP/capita reduced mean statistical distance of EU-15 countries from 1972 to 2004 corresponding to a 10 years moving time window. The line represents the 2-step mobile average fit.

[*] In our used database, the Gross Capital Formation and the Net Exports data are available, for several of the considered countries, until 2003. Therefore, for these two indicators, the last time interval is taken from 2000 to 2003, i.e. for a 4 years time interval.

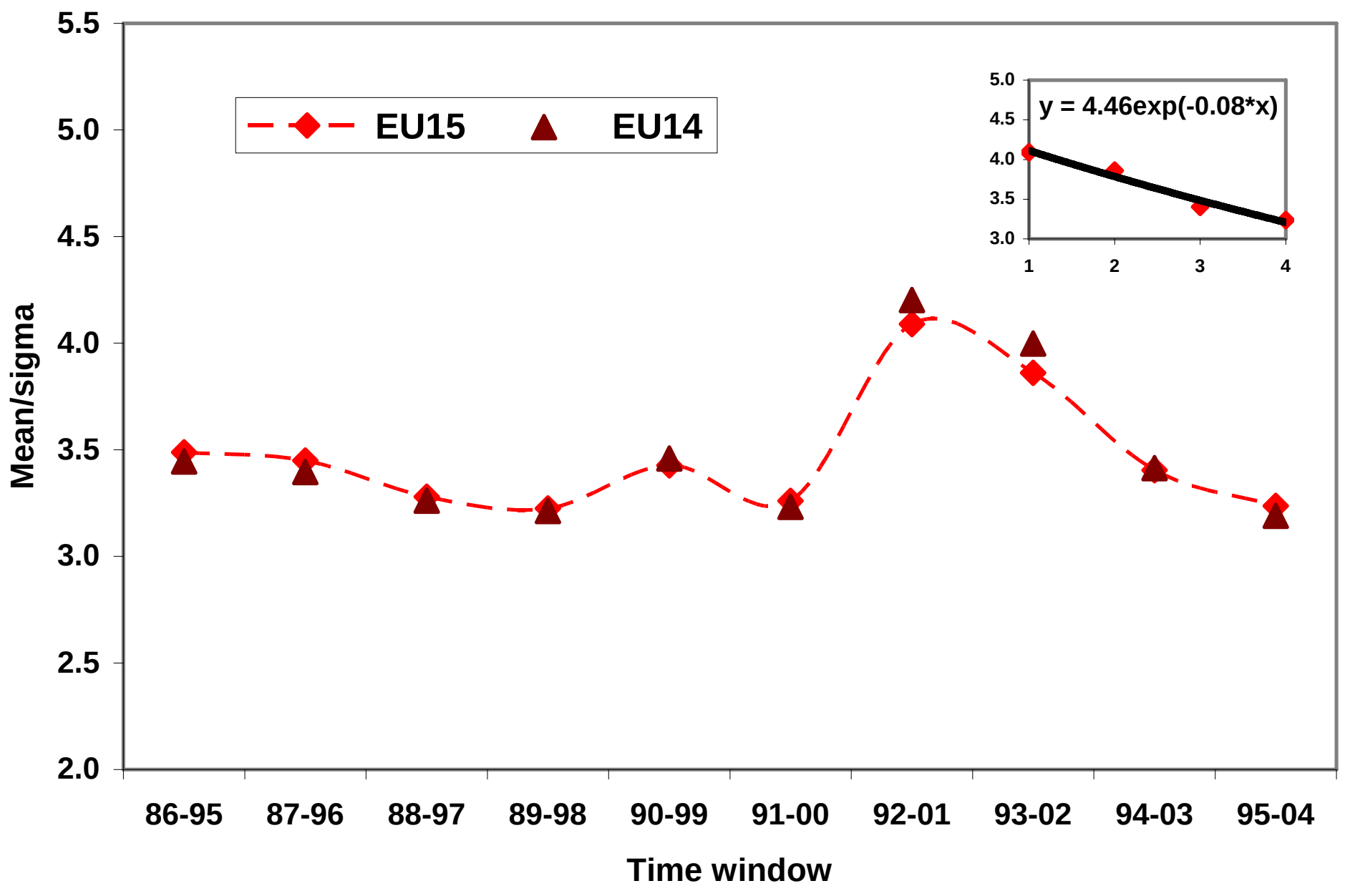

Figure 2: The GDP/capita reduced mean statistical distance of the EU-15 countries (diamond symbol) and EU-14 countries (triangle) respectively (removing DEU), from 1986 to 2004 corresponding to a 10 years moving time window. The inset represents the last 4 points of the main graph, fitted by an exponential. The Pearson RSQ fitting coefficient 0.97.

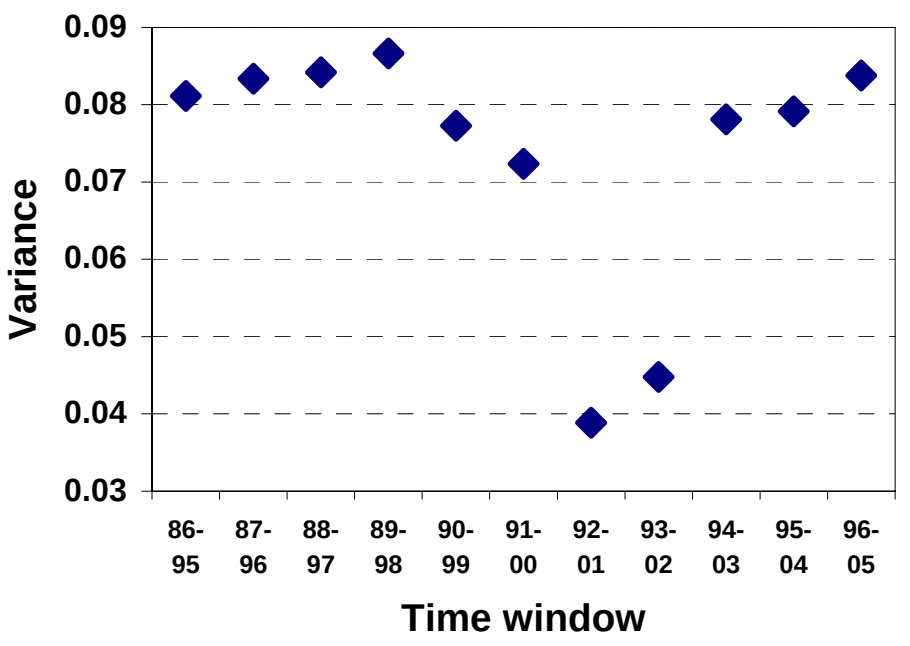

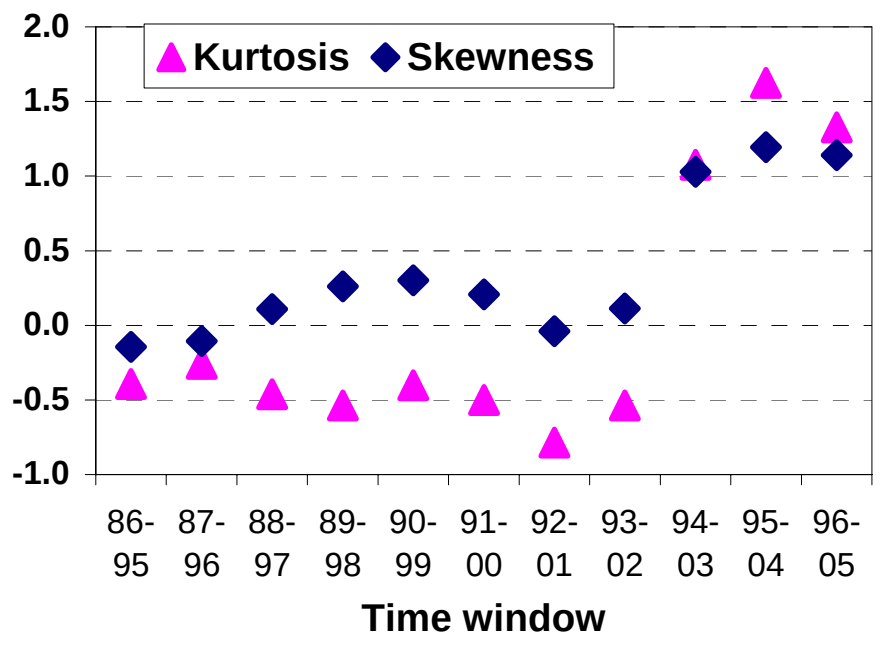

**(a)** **(b)**

Figure 3: Evolution of the common characteristics (variance, kurtosis, skewness) of the distribution of statistical distances in the case of the GDP/capita of EU-15 countries, from 1986 to 2004, shown for a moving 10 years time window.

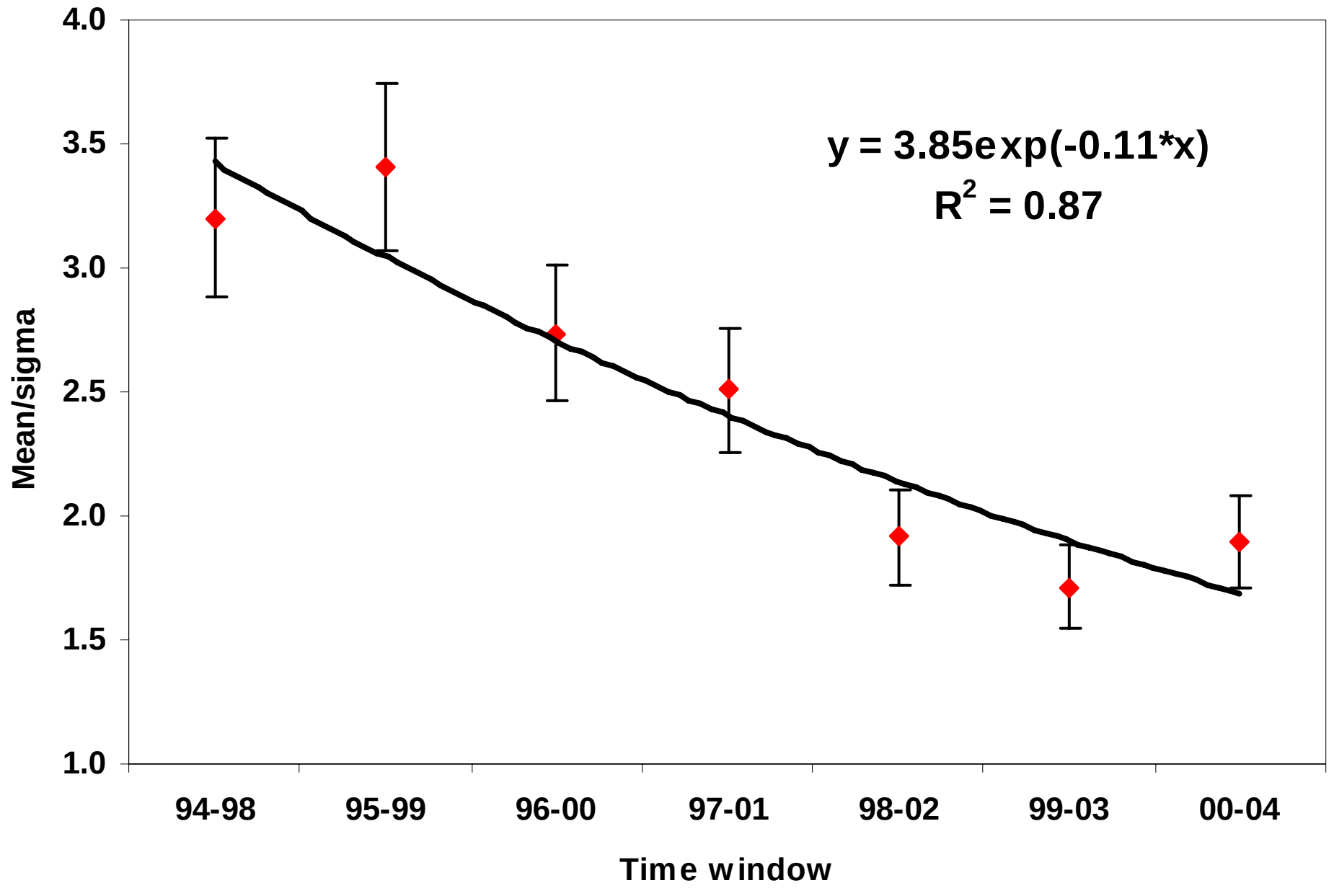

Figure 4: The GDP/capita reduced mean statistical distance of the EU-15 countries from 1994 to 2004 corresponding to a 5 years moving time window. The variable *x* of fit function is the order number of point. $R^2$ is the Pearson RSQ fitting coefficient. Error bars are bootstrap 90% confidence intervals.

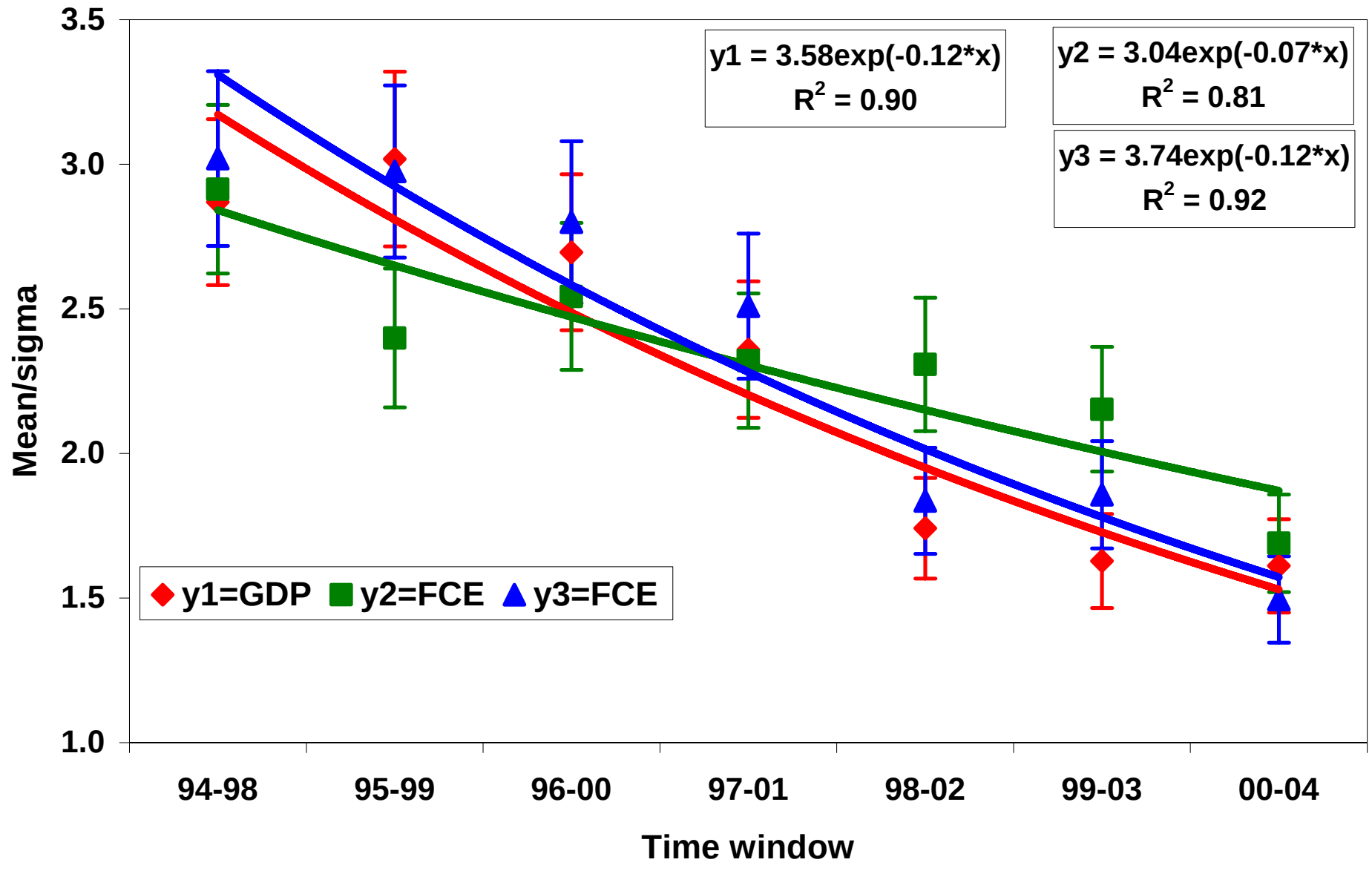

Figure 5: The GDP, FCE and GCF reduced mean statistical distance of the EU-15 countries from 1994 to 2004 corresponding to a 5 years moving time window. The variable *x* of the exponential fit function is the order number of point. $R^2$ is the Pearson RSQ coefficient of fitting. Error bars are bootstrap 90% confidence intervals.

## 3.2 Country clustering structure along the MAMLP method

At this point of our investigation the subsequent ingredients of the MAMLP method, introduced in Sect. 2, are implemented. The first indicator taken into consideration is the GDP annual growth. A virtual ‘AVERAGE’ country is introduced in the system. The statistical distances corresponding to the fixed 5 years moving time window are set in increasing order and the minimal length path (MPL) connections to the AVERAGE are established for each country in every time interval (Table 1a). The resulting hierarchy is found to be changing from a time interval to another. Therefore, another correlation matrix is built, this time for the country movements inside the hierarchy (Table 1b)*. The matrix elements are defined as:

$$\hat{C}_{ij}(t) = \frac{< \hat{d}_i(t)\hat{d}_j(t) > - < \hat{d}_i(t) >< \hat{d}_j(t) >}{\sqrt{< [\hat{d}_i(t)]^2 - < \hat{d}_i(t) >^2 >< [\hat{d}_j(t)]^2 - < \hat{d}_j(t) >^2 >}} \qquad (3.4)$$

where $\hat{d}_i(t)$ and $\hat{d}_j(t)$ are the minimal length path (MPL) distances to the AVERAGE. For simplicity, in Eq. (3.4) are not included the explicit dependencies on the time window size *T*.

In this way the strongest correlations and anti-correlations between GDP fluctuations could be extracted and a clustering structure searched for. We consider that some correlations (and anticorrelations) are more significant than others, as indicated in bold faces, *e.g.* we consider that two countries are strongly tied if: $0.9 \le C \le 1$ or $-1 \le C \le -0.5$ respectively (both intervals enclose about 10% of the total number of the matrix elements). In this sense, a schematic representation of the most significant correlations and anti-correlations taken out from Table 1b can be displayed (Figure 6).

Table 1a: MLP distances to AVERAGE. Indicator: GDP. The moving time window size is 5 years for data taken from 1994 to 2004.

| | AT | BE | DE | DK | ES | FI | FR | UK | GR | IE | IT | LU | NL | PT | SE |
|---|---|---|---|---|---|---|---|---|---|---|---|---|---|---|---|
| **94-98** | .67 | .86 | .86 | .86 | .40 | .40 | .67 | .86 | .40 | .86 | .86 | .40 | .40 | .86 | .86 |
| **95-99** | .60 | .65 | .52 | .71 | .21 | .77 | .45 | .77 | .37 | .65 | .90 | .37 | .23 | .83 | .52 |
| **96-00** | .58 | .32 | .46 | .61 | .34 | .81 | .46 | .32 | .32 | .53 | .32 | .20 | .60 | .60 | .46 |
| **97-01** | .48 | .30 | .48 | .30 | .28 | .42 | .48 | .44 | .68 | .38 | .68 | .14 | .28 | .28 | .48 |
| **98-02** | .43 | .26 | .19 | .19 | .21 | .43 | .19 | .19 | 1.04 | .29 | .44 | .12 | .21 | .21 | .29 |
| **99-03** | .25 | .23 | .19 | .19 | .29 | .26 | .19 | .37 | 1.15 | .26 | .37 | .23 | .19 | .19 | .28 |
| **00-04** | .27 | .27 | .17 | .26 | .28 | .27 | .21 | .27 | .53 | .50 | .28 | .27 | .21 | .21 | .27 |

* For simplicity, all the data enclosed in tables are formatted at two digits after the decimal point

Table 1b: The correlation matrix of country movements inside the hierarchy; Indicator: GDP. The moving time window size is 5 years for data taken from 1994 to 2004.

| | AT | BE | DE | DK | ES | FI | FR | UK | GR | IE | IT | LU | NL | PT | SE |
|---|---|---|---|---|---|---|---|---|---|---|---|---|---|---|---|
| **AT** | 1 | .77 | .88 | .88 | .33 | .69 | .88 | .69 | **-.69** | .75 | .71 | .42 | .61 | .89 | .85 |
| **BE** | | 1 | .88 | **.90** | .41 | .27 | .80 | **.94** | **-.59** | **.92** | .83 | .85 | .23 | **.90** | **.91** |
| **DE** | | | 1 | **.90** | .61 | .35 | **.98** | .86 | **-.65** | .85 | .78 | .61 | .52 | .86 | **.99** |
| **DK** | | | | 1 | .50 | .58 | .87 | .84 | **-.80** | **.93** | .67 | .77 | .58 | **.99** | .88 |
| **ES** | | | | | 1 | -.10 | .61 | .34 | -.38 | .55 | .05 | .36 | .66 | .37 | .64 |
| **FI** | | | | | | 1 | .42 | .25 | **-.62** | .34 | .27 | .14 | .60 | .64 | .26 |
| **FR** | | | | | | | 1 | .79 | **-.71** | .81 | .73 | .52 | .60 | .82 | **.95** |
| **UK** | | | | | | | | 1 | **-.52** | .82 | **.90** | .85 | .12 | .86 | .86 |
| **GR** | | | | | | | | | 1 | **-.82** | -.38 | **-.56** | **-.62** | **-.76** | **-.60** |
| **IE** | | | | | | | | | | 1 | .63 | .85 | .43 | .89 | .87 |
| **IT** | | | | | | | | | | | 1 | .59 | -.05 | .73 | .77 |
| **LU** | | | | | | | | | | | | 1 | .06 | .77 | .65 |
| **NL** | | | | | | | | | | | | | 1. | .50 | .47 |
| **PT** | | | | | | | | | | | | | | 1 | .84 |
| **SE** | | | | | | | | | | | | | | | 1 |

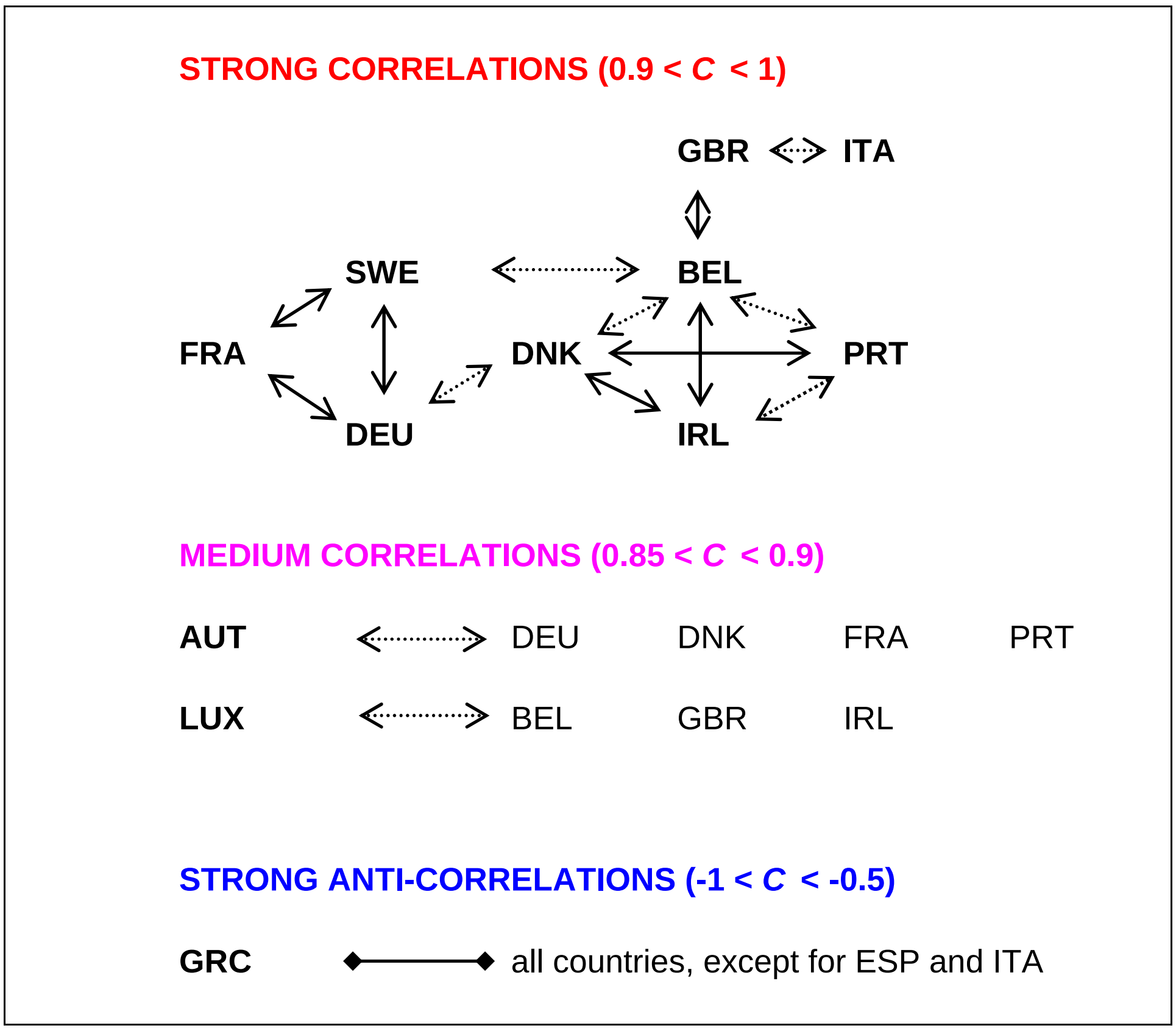

Figure 6: A schematic representation of most significant correlations and anti-correlations taken out from the Table 1b

The MAMLP method can now be applied to the other three macroeconomic indicators defined in Section 2, namely Final Consumption Expenditure, Gross Capital Formation and Net Exports. Tables 2a, 3a and 4a give the corresponding MLP distances to AVERAGE, while Tables 2b, 3b and 4b display the correlation matrices. As for Table 1b, Tables 4b, 5b and 6b display in bold the strongest correlations and anticorrelations.

Table 2a: MLP distances to AVERAGE. Indicator: Final Consumption Expenditure. The moving time window size is 5 years for data taken from 1994 to 2004.

| | AT | BE | DE | DK | ES | FI | FR | UK | GR | IE | IT | LU | NL | PT | SE |
|---|---|---|---|---|---|---|---|---|---|---|---|---|---|---|---|
| **94-98** | .88 | .65 | .85 | .88 | .65 | .37 | .65 | .65 | .65 | .65 | .37 | .65 | .65 | .65 | .65 |
| **95-99** | .79 | .79 | .79 | .81 | .79 | .41 | .79 | .79 | .93 | .79 | .53 | .59 | .79 | .79 | .79 |
| **96-00** | 1.02 | 1.02 | 1.02 | 1.02 | 1.02 | 1.02 | 1.02 | 1.02 | 1.02 | 1.02 | .26 | 1.02 | 1.02 | 1.02 | 1.02 |
| **97-01** | .51 | .51 | .51 | .65 | .51 | .73 | .88 | .51 | .65 | .51 | .33 | .88 | .51 | .51 | .51 |
| **98-02** | .52 | .52 | .52 | .96 | .52 | .66 | .95 | .65 | .96 | .52 | .35 | 1.19 | .52 | .52 | .52 |
| **99-03** | .45 | .42 | .45 | 1.00 | .45 | .53 | .40 | .46 | 1.00 | .42 | .30 | .92 | .45 | .45 | .45 |
| **00-04** | .88 | .65 | .85 | .88 | .65 | .37 | .65 | .65 | .65 | .65 | .37 | .65 | .65 | .65 | .65 |

Table 2b: The correlation matrix of country movements inside the hierarchy. Indicator: Final Consumption Expenditure. The moving time window size is 5 years for data taken from 1994 to 2004.

| | AT | BE | DE | DK | ES | FI | FR | UK | GR | IE | IT | LU | NL | PT | SE |
|---|---|---|---|---|---|---|---|---|---|---|---|---|---|---|---|
| **AT** | 1 | **.92** | **1** | .23 | **.92** | .21 | .38 | .87 | .03 | **.92** | .07 | -.34 | **.92** | **.92** | **.92** |
| **BE** | | 1 | **.94** | .23 | **1** | .45 | .56 | **.97** | .28 | **1** | .06 | -.15 | **1** | **1** | **1** |
| **DE** | | | 1 | .24 | **.93** | .24 | .40 | .89 | .07 | **.94** | .07 | -.32 | **.93** | **.93** | **.93** |
| **DK** | | | | 1 | .26 | .22 | -.14 | .35 | .75 | .23 | -.41 | .44 | .26 | .26 | .26 |
| **ES** | | | | | 1 | .45 | .53 | **.97** | .31 | **1** | .04 | -.15 | **1** | **1** | **1** |
| **FI** | | | | | | 1 | .65 | .49 | .34 | .45 | **-.68** | .68 | .45 | .45 | .45 |
| **FR** | | | | | | | 1 | .64 | .05 | .56 | -.05 | .38 | .53 | .53 | .53 |
| **UK** | | | | | | | | 1 | .40 | **.97** | .03 | .02 | **.97** | **.97** | **.97** |
| **GR** | | | | | | | | | 1 | .28 | -.11 | .45 | .31 | .31 | .31 |
| **IE** | | | | | | | | | | 1 | .06 | -.15 | **1** | **1** | **1** |
| **IT** | | | | | | | | | | | 1 | **-.68** | .04 | .04 | .04 |
| **LU** | | | | | | | | | | | | 1 | -.15 | -.15 | -.15 |
| **NL** | | | | | | | | | | | | | 1 | **1** | **1** |
| **PT** | | | | | | | | | | | | | | 1 | **1** |
| **SE** | | | | | | | | | | | | | | | 1 |

Table 3a: MLP distances to AVERAGE. Indicator: Gross Capital Formation. The moving time window size is 5 years for data taken from 1994 to 2003.

| | AT | BE | DE | DK | ES | FI | FR | UK | GR | IE | IT | LU | NL | PT | SE |
|---|---|---|---|---|---|---|---|---|---|---|---|---|---|---|---|
| **94-98** | .51 | .48 | .59 | .52 | .66 | .48 | .66 | .58 | .89 | .67 | .38 | .85 | .67 | .37 | .51 |
| **95-99** | .47 | .46 | .75 | .49 | .54 | .46 | .54 | .61 | .75 | .49 | .33 | .83 | .49 | .39 | .58 |
| **96-00** | .75 | .78 | .75 | .78 | .75 | .78 | .75 | .58 | .75 | .84 | .32 | .32 | .48 | .20 | .75 |
| **97-01** | .70 | .47 | .70 | .62 | .70 | .62 | .70 | .57 | .70 | .38 | .63 | .29 | .29 | .09 | .70 |
| **98-02** | .46 | .46 | .46 | .68 | .46 | .68 | .46 | .61 | .46 | .46 | 1.13 | .46 | .46 | .46 | .46 |
| **99-03** | .70 | .70 | .70 | .88 | .70 | .88 | .70 | .70 | .70 | .70 | 1.07 | .70 | .70 | .70 | .70 |

Table 3b: The correlation matrix of country movements inside the hierarchy. Indicator: Gross Capital Formation. The moving time window size is 5 years for data taken from 1994 to 2003.

| | AT | BE | DE | DK | ES | FI | FR | UK | GR | IE | IT | LU | NL | PT | SE |
|---|---|---|---|---|---|---|---|---|---|---|---|---|---|---|---|
| **AT** | 1 | .76 | .59 | .68 | .88 | .69 | .88 | .10 | .19 | .45 | -.04 | -.58 | -.12 | -.26 | **.94** |
| **BE** | | 1 | .47 | .81 | .67 | .79 | .67 | .35 | .15 | .85 | -.02 | -.27 | .32 | .15 | .73 |
| **DE** | | | 1 | .10 | .64 | .09 | .64 | .05 | .55 | .30 | -.57 | -.02 | -.08 | -.25 | .81 |
| **DK** | | | | 1 | .41 | **1** | .41 | .61 | -.32 | .50 | .56 | -.40 | .24 | .39 | .55 |
| **ES** | | | | | 1 | .40 | **1** | -.04 | .61 | .58 | -.35 | -.26 | .11 | -.29 | .83 |
| **FI** | | | | | | 1 | .40 | .58 | -.37 | .46 | .57 | -.46 | .17 | .35 | .56 |
| **FR** | | | | | | | 1 | -.04 | .61 | .58 | -.35 | -.26 | .11 | -.29 | .83 |
| **UK** | | | | | | | | 1 | -.21 | .20 | .63 | .37 | .61 | **.91** | .12 |
| **GR** | | | | | | | | | 1 | .44 | -.76 | .45 | .37 | -.20 | .27 |
| **IE** | | | | | | | | | | 1 | -.26 | .10 | .62 | .21 | .40 |
| **IT** | | | | | | | | | | | 1 | -.15 | .12 | .60 | -.21 |
| **LU** | | | | | | | | | | | | 1 | .73 | .60 | -.46 |
| **NL** | | | | | | | | | | | | | 1 | .78 | -.17 |
| **PT** | | | | | | | | | | | | | | 1 | -.27 |
| **SE** | | | | | | | | | | | | | | | 1 |

Table 4a: MLP distances to AVERAGE. Indicator: Net Exports. The moving time window size is 5 years for data taken from 1994 to 2003.

| | AT | BE | DE | DK | ES | FI | FR | UK | GR | IE | IT | LU | NL | PT | SE |
|---|---|---|---|---|---|---|---|---|---|---|---|---|---|---|---|
| **94-98** | 1.27 | .19 | .65 | .89 | .45 | .80 | .65 | .62 | .75 | .62 | .62 | .80 | .64 | .62 | .62 |
| **95-99** | 1.13 | .40 | .66 | 1.11 | .66 | .87 | .66 | .56 | .87 | .56 | .56 | .87 | 1.11 | .56 | .56 |
| **96-00** | 1.29 | .72 | .52 | .81 | .52 | .81 | .56 | .22 | .81 | .72 | .54 | .81 | .54 | .54 | .72 |
| **97-01** | 1.06 | .55 | .64 | .80 | .64 | .70 | .64 | .26 | .39 | .55 | .64 | .70 | .64 | .64 | .55 |
| **98-02** | .94 | .73 | .54 | .73 | .54 | .67 | .73 | .54 | .54 | .73 | .54 | .67 | .67 | .54 | .73 |
| **99-03** | .37 | .65 | .37 | 1.03 | .50 | .82 | .79 | .76 | .65 | .79 | .50 | .82 | .82 | .37 | .79 |

Table 4b: The correlation matrix of country movements inside the hierarchy. Indicator: Net Exports. The time moving window size is 5 years for data taken from 1994 to 2003.

| | **AT** | **BE** | **DE** | **DK** | **ES** | **FI** | **FR** | **UK** | **GR** | **IE** | **IT** | **LU** | **NL** | **PT** | **SE** |
|---|---|---|---|---|---|---|---|---|---|---|---|---|---|---|---|
| **AT** | 1 | -.39 | .80 | -.32 | .11 | .02 | **-.89** | -.62 | .30 | **-.59** | .60 | .02 | -.26 | .84 | **-.59** |
| **BE** | | 1 | **-.65** | -.39 | .09 | -.39 | .15 | -.30 | -.32 | .62 | **-.61** | -.39 | -.27 | -.48 | .62 |
| **DE** | | | 1 | -.07 | .44 | -.05 | -.56 | -.35 | .06 | -.92 | .82 | -.05 | .13 | **.93** | **-.92** |
| **DK** | | | | 1 | .22 | .85 | .28 | .56 | .58 | -.14 | -.28 | .85 | .86 | -.41 | -.14 |
| **ES** | | | | | 1 | -.03 | -.16 | -.37 | -.18 | **-.64** | .23 | -.03 | .53 | .30 | **-.64** |
| **FI** | | | | | | 1 | -.13 | .30 | .86 | -.04 | -.29 | **1** | .56 | -.31 | -.04 |
| **FR** | | | | | | | 1 | .82 | -.29 | .47 | -.47 | -.13 | .35 | **-.67** | .47 |
| **UK** | | | | | | | | 1 | .21 | .34 | -.40 | .30 | .50 | **-.57** | .34 |
| **GR** | | | | | | | | | 1 | .05 | -.35 | .86 | .40 | -.16 | .05 |
| **IE** | | | | | | | | | | 1 | **-.82** | -.04 | -.28 | **-.81** | **1** |
| **IT** | | | | | | | | | | | 1 | -.29 | -.24 | **.90** | **-.82** |
| **LU** | | | | | | | | | | | | 1 | .56 | -.31 | -.04 |
| **NL** | | | | | | | | | | | | | 1 | -.25 | -.28 |
| **PT** | | | | | | | | | | | | | | 1 | **-.81** |
| **SE** | | | | | | | | | | | | | | | 1 |

In the above tables we can observe the position of the bold elements, whence see that five of the mostly correlated countries with respect to GDP fluctuations (SWE-GBR-DEU-BEL-IRL) also display strong correlations in the Final Consumption Expenditure and medium correlations in Gross Capital Formation fluctuations ($C_{ij} \sim 0.8$). Moreover, some of them are strongly anticorrelated in Net Exports fluctuations (e.g. $C_{ij} < -0.9$ for DEU-SWE and DEU-IRL). The top strong correlations appear in FCE fluctuations (Table 2b), while the top anticorrelations can be found in NEX fluctuations (Table 4b).

Finally, we calculate a so called *sensitivity degree*, i.e., the quadratic sum of all the correlation coefficients:

$$(\chi_i)_\alpha = \sum_{\substack{i,j=1 \\ i \neq j}}^{N} (\hat{C}_{ij})^2 \tag{3.5}$$

where $\alpha \equiv$ GDP, FCE, GCF and NEX. The results are given in Table 5 for all considered indicators and for each country.

Table 5: The quadratic sum of correlation coefficients (the *sensitivity degree* of countries) for the fluctuations of GDP, Final Consumption Expenditure (FCE), Gross Capital Formation (GCF) and Net Exports (NEX), for data taken from 1994 to 2004 (GDP and FCE) and from 1994 to 2003 (GCF and NEX) respectively..

| GDP | | FCE | | GCF | | NEX | |
|---|---|---|---|---|---|---|---|
| **DK** | 9.08 | **BE** | 8.34 | **AT** | 4.99 | **PT** | 5.23 |
| **PT** | 8.71 | **IE** | 8.34 | **SE** | 4.69 | **DE** | 4.92 |
| **DE** | 8.68 | **ES** | 8.32 | **ES** | 4.66 | **IE** | 4.76 |
| **SE** | 8.47 | **NL** | 8.32 | **FR** | 4.66 | **SE** | 4.76 |
| **IE** | 8.26 | **PT** | 8.32 | **BE** | 4.58 | **IT** | 4.41 |
| **BE** | 8.25 | **SE** | 8.32 | **DK** | 4.18 | **AT** | 3.99 |
| **FR** | 8.21 | **UK** | 8.14 | **FI** | 4.09 | **DK** | 3.50 |
| **AT** | 7.60 | **DE** | 7.42 | **IE** | 3.04 | **FR** | 3.24 |
| **UK** | 7.59 | **AT** | 7.15 | **PT** | 2.89 | **FI** | 3.23 |
| **IT** | 5.68 | **FR** | 3.07 | **DE** | 2.85 | **LU** | 3.23 |
| **GR** | 5.64 | **FI** | 3.06 | **IT** | 2.70 | **UK** | 2.91 |
| **LU** | 5.40 | **LU** | 1.81 | **UK** | 2.68 | **BE** | 2.71 |
| **NL** | 3.25 | **DK** | 1.61 | **GR** | 2.63 | **NL** | 2.63 |
| **ES** | 2.97 | **GR** | 1.60 | **LU** | 2.39 | **GR** | 2.49 |
| **FI** | 2.68 | **IT** | 1.13 | **NL** | 2.31 | **ES** | 1.69 |

# 4 Conclusion

Analyzing the time evolution of the mean statistical distance between the EU-15 countries one expects to find a decreasing trend, when one expects a so called economy globalization. For the 10 years moving time window size (Figures 1 and 3) one can see a decreasing trend between 1979 and 1992 and for the last 4 time intervals, *i.e.*, the period 1992-2004, when the mean distance decreases from 4.80 to 3.20 and from 4.09 to 3.06 respectively (in $m/\sigma$ units, where $m$ = the mean and $\sigma$ = the standard deviation). In return, taking into account the whole evolution, the phenomenon appears as strongly nonlinear and non-monotonic. A somewhat unexpected evolution is registered in 1991-2000 and 1992-2001, when the mean distance abruptly increases (in a single step) from 3.26 to 4.09. It is not only a change of value but also a change of trend (Figure 2), i.e., from a quasi-constant trend (or a slow linear decrease) to another one that is strongly decreasing well fitted to an exponential. The abrupt change of trend also occurred for other statistical parameters of the distance distributions, *e.g.* the variance, kurtosis and skewness (Figure 3), approximately in the same time interval or in the next one.

The first explanation one could imagine would be the Berlin Wall fall and Germany re-unification. Indeed, Germany was taken into consideration in the previous estimation of the mean distance and by far, it was having the most abrupt variation of economic parameters in that period (see e.g. Keller, 1997). But the phenomenon seems to be somewhat more complex. In Figures 1a and 1b it has been seen that the time variation of the mean distance between countries with or without Germany (and its connections) (the EU-14 plot), is not at all affected.

Another explanation might be found when analyzing several other important events which occurred *after* the Berlin wall fall *i.e.* the political changes and opening of new markets in Eastern Europe and Central Asia, while the Western European countries and their investors were having different positions in relation with these new possibilities of investment. In physical terms one can say that there was an increase in ‘physical volume or available space for the economic gas’ or, in other words a diffusion process as described in ACP model (Ausloos et al., 2004), which generated an abrupt increase of the mean distance between countries. It is interesting to note that in physical models these nonequilibrium abrupt transitions, due to ‘’shocks”, are generally followed by exponential or power law relaxations, (Lambiotte and Ausloos, 2006; Sornette et al., 2004).

As results from the above considerations, in large time windows (*e.g.* 10 years or more) some successive increase and decrease of the mean statistical distance arise, so that the globalization trend appears to be non-monotonic. This aspect is somewhat expected, taking into account the large variations in the economic/social/political environment.

On the contrary, when a 5 years time window size is moved over the interval 1994-2004, there is a clear decrease of the mean statistical distance between EU-15 countries from 3.20 to 1.89 as concerns GDP/capita (Figure 4), from 2.86 to 1.81 for GDP, from 2.91 to 1.68 for the Final Consumption and from 3.01 to 1.49 for the Capital Growth (Figure 5). The mean distance does not display a clear trend as regards Net Exports fluctuations – at least in this time window size.

Regarding the country clusters, as in other classification problems, a major issue that arises when the classification trees derive from real data with much random noise concerns how to define what a cluster is. This general issue is discussed in the literature on tree classification under the topic of *overfitting* (Breiman et al., 1984) If not stopped, the tree algorithm will ultimately “extract” all information from the data, including random or noise variation.

To avoid this trap, in our classification we have considered as “strong” correlations and anti-correlations those with $C \geq 0.9$ and $C \leq -0.5$ respectively, taking into account that the both intervals of $C$ include the same percentage (~ 10 %) from the total set of correlation coefficients. From this criterion, the strongly correlated countries in GDP fluctuations (Table 1b and Figure 6) can be partitioned into two clusters: FRA-SWE-DEU and BEL-GBR-IRE-DNK-PRT. ITA can be considered in the second cluster for its strong correlation with GBR, but it does not display any strong correlations with the other countries. LUX is weakly correlated to the second cluster, while AUT is somewhat “equidistant” displaying medium correlations with both clusters. GRC holds a special position: its GDP fluctuations appear to be strongly anti-correlated with of all other countries.

The sum of squares of the correlation coefficients (Table 5) measures the country so called “sensitivity degree”. The most sensitive countries to the GDP fluctuations of others seem to be DNK, PRT, DEU, SWE and IRL. The least “sensitive” ones are NLD, ESP and FIN. Geographically closed pairs of countries (FIN-DNK, ESP-PRT and NLD-DEU) are situated at opposite ends of this classification. We stress as an original finding that the sensitivity degree is not necessary related to the geographical position, and thus to the most probable trade exchanges.

The strongest correlations of GDP fluctuations can be recovered only among the strongest correlations of the Final Consumption index. Five of the mostly correlated countries in GDP fluctuations (SWE-GBR-DEU-BEL-IRL) appear as such (Table 1b), due their strong correlation in the Final Consumption fluctuations (Table 2b). They display medium correlations in Gross Capital Formation fluctuations ($C_{ij} \sim 0.8$) and some of them are even strongly anticorrelated in Net Exports fluctuations (*e.g.* $C_{ij} < -0.9$ for DEU-SWE and DEU-IRL). On the other hand, one can note that the sensitivity classifications regarding GDP and Net Exports fluctuations are quite similar (Table 5), at least for the countries situated at the top and at the bottom. We recover here one of the main characteristics of social networks that is the positive correlation existing between the node degrees (Ramasco et al., 2003), *i.e.* the high connected countries commonly tend to connect with other well connected ones. So, a true proof of globalization is found.

Finally, we recall that the above classifications and clusters refer only to the fluctuations of investigated quantities, without consideration of their actual value. It requires some further work to observe whether our results are carried over to other correlation-hierarchical clustering of EU countries in other situations.

# Colophon

*Acknowledgements:* Mircea Gligor thanks the Franqui Foundation for a fellowship allowing him a stay in Liège. Marcel Ausloos thanks ARC for some financial support. Part of this work has been presented at the COST P10 Vilnius meeting in May 2006. Both authors thank such a European program for some support.

# Appendix: Shuffled data analysis

For a robustness test and statistical error bar significance, the elements of the statistical distance matrices were shuffled per columns so as the data proceeded from different time windows were randomly mixed. The mean statistical distances so constructed are plotted in figures 7a (for GDP), 7b (for FCE) and 7c (for GCF). For comparison, the true data and their best fit are shown as well.

In all three index cases so considered, the mean distance derived from the shuffled data midly oscillates around a constant value, as it has to be expected; the amplitude of the fluctuations is 0.49 units mean/sigma for GDP, 0.12 units for FCE and 0.28 units for GCF, that means 35 %, 9.7 %, and 21.5 % respectively from their maximal (real) variation induced by the decreasing trend.

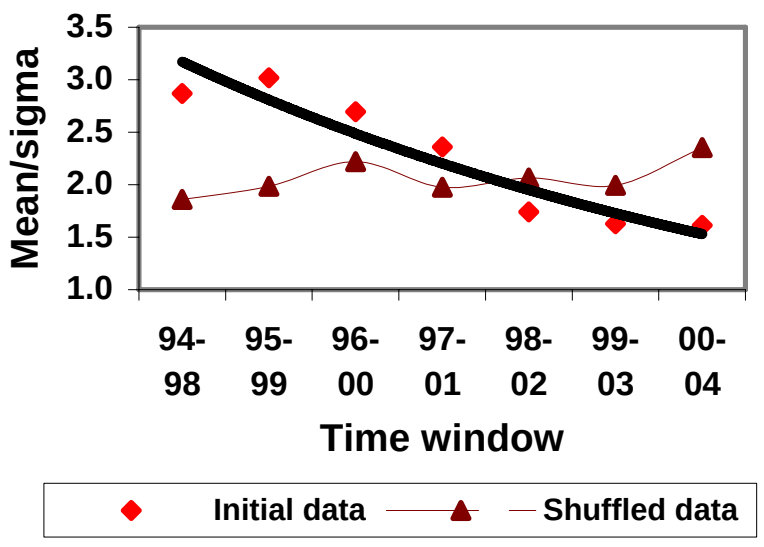

(a)

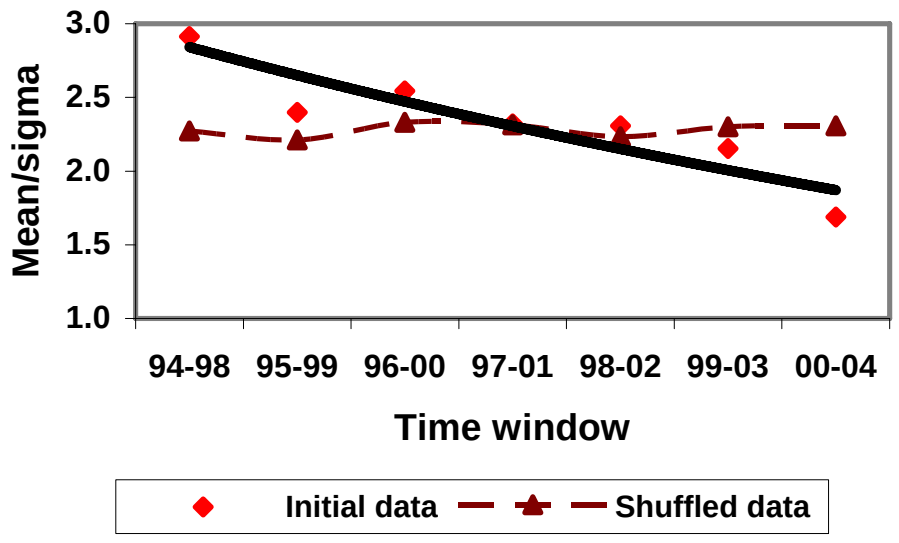

(b)

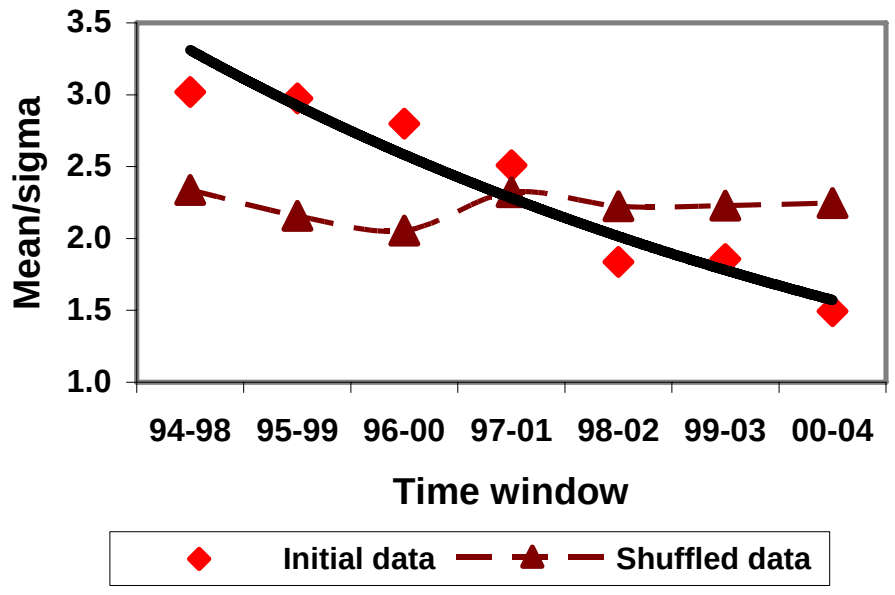

(c)

Figure 7: The reduced mean distance between countries regarding to (a) GDP; (b) Final Consumption Expenditure; (c) Gross Capital Formation fluctuations, derived from shuffled (triangles) and initial (diamond symbol) datasets. The continuous line is the exponential fit to the initial data.

As a second test, the correlation matrix from Table 1b was randomized by shuffling MLP distances to AVERAGE (from Table 1a), firstly per columns and secondly per lines. The results are presented in Table 6. The maximum and minimum values of the correlation coefficients are found to be $(C_{max})_{\text{shufll}}$ = 0.71 and $(C_{min})_{\text{shufll}}$ = – 0.68 as compared with $(C_{max})$ = 0.99 and $(C_{min})$ = – 0.80 from Table 1b. According to the criterion discussed in Section 5 ($C_{corr} \geq 0.9$ and $C_{anticorr} \leq -0.8$), one can say that neither any strong correlations nor anti-correlation appear. In other words, the correlations which resulted in the clustering structure from Figure 6 are destroyed by the randomization, consequently giving weight to the main text results, analysis and conclusion.

Table 6: The randomized correlation matrix of country movements of inside the hierarchy. Indicator: GDP. Time window size: 5 years

| | **AT** | **BE** | **DE** | **DK** | **ES** | **FI** | **FR** | **UK** | **GR** | **IE** | **IT** | **LU** | **NL** | **PT** | **SE** |
|---|---|---|---|---|---|---|---|---|---|---|---|---|---|---|---|
| **AT** | 1 | .19 | -.07 | -.28 | .23 | -.23 | .45 | .55 | -.47 | .07 | -.35 | .28 | -.43 | .29 | -.49 |
| **BE** | | 1 | .51 | .10 | -.10 | -.47 | .16 | .24 | -.35 | -.48 | -.61 | .41 | .07 | -.55 | .18 |
| **DE** | | | 1 | .53 | .24 | -.22 | **.70** | -.22 | -.48 | -.50 | -.11 | -.34 | -.02 | .24 | .16 |
| **DK** | | | | 1 | -.32 | .19 | .19 | .27 | -.20 | -.64 | -.22 | -.67 | -.15 | .36 | .34 |
| **ES** | | | | | 1 | .42 | .58 | -.57 | -.60 | .32 | .66 | -.21 | .06 | .37 | .15 |
| **FI** | | | | | | 1 | .00 | -.16 | -.17 | -.02 | **.71** | -.67 | .28 | .33 | .43 |
| **FR** | | | | | | | 1 | -.06 | -.53 | -.33 | .17 | -.44 | .00 | .62 | -.32 |
| **UK** | | | | | | | | 1 | .00 | -.46 | **-.68** | .09 | -.23 | .00 | -.32 |
| **GR** | | | | | | | | | 1 | -.05 | .08 | .10 | .50 | -.37 | -.42 |
| **IE** | | | | | | | | | | 1 | .26 | .44 | -.44 | .05 | .08 |
| **IT** | | | | | | | | | | | 1 | -.52 | .47 | .32 | .10 |
| **LU** | | | | | | | | | | | | 1 | -.22 | **-.67** | -.12 |
| **NL** | | | | | | | | | | | | | 1 | -.40 | -.12 |
| **PT** | | | | | | | | | | | | | | 1 | -.21 |
| **SE** | | | | | | | | | | | | | | | 1 |